\documentstyle[12pt,aps,prc,tighten]{revtex}

\newcommand{\greac}{\bar{p}p\rightarrow\bar{Y}Y}
\newcommand{\reac}{\bar{p}p\rightarrow\bar{\Lambda}\Lambda}
\newcommand{\sigv}{{\vec \sigma}}
\newcommand{\lh}{\bf \hat{l}}
\newcommand{\mh}{\bf \hat{m}}
\newcommand{\nh}{\bf \hat{n}}
\newcommand{\ph}{\bf \hat{p}}
\newcommand{\xh}{\bf \hat{x}}
\newcommand{\yh}{\bf \hat{y}}
\newcommand{\zh}{\bf \hat{z}}
\newcommand{\cx}{{\cal X}}

\begin{document}

\title{Spin Observables in $\reac$ with a Transverse Initial State Polarization}
\author{Kent Paschke and Brian Quinn}

\address{Carnegie Mellon University\\ Pittsburgh, PA 15217}
\maketitle

\begin{abstract}
The formalism describing the scattering of two spin-$\frac{1}{2}$ 
objects is reviewed for the case of $\reac$. It is shown that an 
experiment utilizing a transverse target polarization can, in principle, 
completely determine the spin structure of the reaction.  Additional 
measurements, even those using both beam and target polarizations,
would not be sensitive to any additional spin dynamics.  Thus, the 
transverse target polarization allows access to the complete set of 
spin observables, not just the subset upon which the literature has 
previously focused. This discussion is especially relevant in light 
of the data collected by PS185/3 at LEAR.
\end{abstract}

\vskip .5cm

Exclusive production of antihyperon-hyperon states from antiproton-proton 
annihilation ($\bar{p}p\rightarrow\overline{Y}Y$) has been extensively studied near threshold by the 
PS185 collaboration throughout the 13 year lifetime of LEAR at 
CERN~\cite{PS185M,PS185_96}. Results for the total and differential 
cross-sections and those spin observables depending only on final state 
spins have been published for a range of energies and channels.  In 1996, 
an upgraded version of the experiment, PS185/3, incorporated a transverse 
target polarization to allow 
sensitivity to a wider range of spin observables in the reaction 
$\reac$~\cite{PS185_3}.

There is significant interest in the results of this measurement.  Two types of
models with different physical motivations have been successfully used to describe 
the reaction in this energy regime: $K$ and $K^{\ast}$ meson 
exchange~\cite{Haidenbauer,Meson} and effective quark 
models~\cite{Alberg93,Quark}. While careful applications of each of these 
approaches have been able to describe the existing data with reasonable accuracy, 
differing predictions are obtained for those observables which depend on initial 
state spin~\cite{Haidenbauer,Alberg93}. There has also been growing interest in 
the relevance of the spin structure of $\reac$ to the question of the strangeness 
component of the nucleon~\cite{Alberg95,Pak99}.

The formalism for scattering two spin-$\frac{1}{2}$ particles to two
spin-$\frac{1}{2}$ particles has been well developed in both the general case
and in application to the $\greac$ reaction
~\cite{Wolfenstein,Bystricky78,LaFrance80,Durand}. 
The implications of this application of the formalism must be 
carefully explored for the larger set of observables available 
to PS185/3~\cite{Richard1,Richard2}.  The recent
literature discussing the strangeness production dynamics has 
focused on the depolarization $D_{nn}$, and, to a lesser extent, 
other rank-2 spin observables to which PS185/3 is directly 
sensitive~\cite{PS185_3,Haidenbauer,Alberg93,Alberg95,Richard1}.  As will be 
shown below, the use of a transverse target polarization, in principle, 
allows the complete determination of the spin structure of the reaction. 
This then permits model independent predictions of the full set of spin 
observables, including those that were not directly measured. 
This is the first demonstration of the 
feasibility of such an analysis for $\reac$, although 
this possibility had been hinted at by previous publications~\cite{Durand}.

In discussing spin-$\frac{1}{2}$ particle scattering it is common to 
parameterize the spin structure of the transition matrix ${\cal M}$ explicitly.
For $\reac$, the transition matrix ${\cal M}$ can be written in the center of 
mass frame as a function of 6 complex amplitudes 
$\{a,b,c,d,e,g\}$~\cite{Haidenbauer,LaFrance80}:
\begin{equation}\label{TransMat}
{\cal M}_{\bar{p}p\rightarrow\bar{\Lambda}\Lambda} =
\frac{1}{2} \left[ \begin{array}{c}
  (a+b) + (a-b)\sigv_{1}\cdot\nh\,\sigv_{2}\cdot\nh 
+ (c+d)\sigv_{1}\cdot\mh\,\sigv_{2}\cdot\mh \\
+ (c-d)\sigv_{1}\cdot\lh\,\sigv_{2}\cdot\lh 
+ e(\sigv_{1}\cdot\nh + \sigv_{2}\cdot\nh ) \\
+ g(\sigv_{1}\cdot\lh\,\sigv_{2}\cdot\mh+
\sigv_{1}\cdot\mh\,\sigv_{2}\cdot\lh)  
\end{array} \right] \end{equation}
The unit vectors are defined so that $\lh$ and $\mh$ lie in the scattering plane, 
and $\nh$ is the normal to the scattering plane.  $\sigv_{1}$ represents the
vector of Pauli spin matrices for the antiproton in the initial state and the 
$\bar{\Lambda}$ in the final state, while $\sigv_{2}$ represents these 
matrices for the proton and $\Lambda$.  This expression 
embodies the parity and C-parity symmetries of the strong interaction 
but is otherwise entirely model independent.  The spin observables are given by
\begin{equation}\label{Def_sp_obs}
\cx_{ij\mu\nu} = \frac{\frac{1}{4}Tr\left[
\sigma_{1i}\sigma_{2j}{\cal M}\sigma_{1\mu}\sigma_{2\nu}{\cal M}^{\dagger}
\right]}{I_{0}}   \end{equation}
Here each index can take any value from the set $\{0,l,m,n\}$ with $\sigma_{0}$ 
defined as the identity operator.  
$I_{0}=\frac{1}{4}Tr\left[{\cal M}{\cal M}^{\dagger}\right]$ 
represents the differential cross-section averaged over the spin of the particles.
Measurements of the spin observables with non-zero first and second indices 
would require a measurement of the final state spins, while measurements of the 
observables with non-zero third or fourth indices would require the measurement
of the angular distribution relative to a polarization of the beam or target, 
respectively.

The self analyzing weak decay of the hyperons allows access to the $\cx_{ij00}$
terms from measurements of the angular distribution of the decay products of 
the hyperon and antihyperon.    The use of a transversely polarized target 
introduces a dependence on the angle of the scattering plane relative to the 
target polarization. This allows measurement of all observables 
$\cx_{ij0\nu}$ with the index $\nu$ restricted by the direction of target 
polarization. In this case, a convenient choice of axes is
\begin{equation} \label{axes} 
\left\{ \nh = \frac{\ph_{1{\it i}}\times\ph_{1{\it f}}}
         {|\ph_{1{\it i}}\times\ph_{1{\it f}}|} \:\:,\:\:
\lh = \ph_{1i} \:\: , \:\: \mh = \nh\times\lh \right\} \end{equation}
where $\ph_{1i}$ is the direction of the incident anti-proton and  
$\ph_{1f}$ is the direction of the produced anti-lambda.  

With this choice, the  index $\nu$ is restricted to $\{0,m,n\}$ for
a transverse target polarization.  Of those 48 
available observables $\cx_{ij0\nu}$, 24 are trivially zero by parity 
considerations.  Explicitly, we have:
\begin{equation} \label{relations} \begin{array}{llclcl}
I_{0} & & = & \frac{1}{2} \{ |a|^{2} + |b|^{2} + |c|^{2} + |d|^{2} + 
|e|^{2} + |g|^{2} \} & & \\
I_{0} C_{nn00} & & = & \frac{1}{2} \{ |a|^{2} - |b|^{2} - |c|^{2} + |d|^{2} + 
|e|^{2} + |g|^{2} \} & = & I_{0} C_{yy} \\
I_{0} D_{0n0n} & & = & \frac{1}{2} \{ |a|^{2} + |b|^{2} - |c|^{2} - |d|^{2} + 
|e|^{2} - |g|^{2} \} & = & I_{0} D_{yy} \\
I_{0} K_{n00n} & & = & \frac{1}{2} \{ |a|^{2} - |b|^{2} + |c|^{2} - |d|^{2} + 
|e|^{2} - |g|^{2} \} & = & I_{0} K_{yy} \\
I_{0} P_{0n00} & = I_{0} P_{n000} 
               & = &  Re(a^{\ast}e) - Im(d^{\ast}g) & = & I_{0}P_{y}\\
I_{0} A_{000n} & = I_{0} C_{nn0n} 
               & = &  Re(a^{\ast}e) + Im(d^{\ast}g) & = & I_{0}A_{y}\\
I_{0} C_{ml00} & = I_{0} C_{lm00} 
               & = &  Re(a^{\ast}g) + Im(d^{\ast}e) & = & -I_{0}C_{xz}\\
I_{0} C_{mm0n} & = -I_{0} C_{ll0n} 
               & = & Re(d^{\ast}e) + Im(a^{\ast}g)  & = & -I_{0}C_{xx0y}\\
I_{0} C_{mm00} & & =
               & Re( a^{\ast}d + b^{\ast}c) + Im(e^{\ast}g)& = &-I_{0}C_{xx}\\
I_{0} C_{ll00} & & = 
               & Re(-a^{\ast}d + b^{\ast}c) - Im(e^{\ast}g)& = &-I_{0}C_{zz}\\
I_{0} C_{lm0n} & & = 
               & Re(e^{\ast}g) + Im(-a^{\ast}d + b^{\ast}c) & = &-I_{0}C_{zx0y}\\
I_{0} C_{ml0n} & & = 
               & Re(e^{\ast}g) + Im(-a^{\ast}d - b^{\ast}c) & = &-I_{0}C_{xz0y}\\
I_{0} D_{0m0m} & & = 
               & Re(a^{\ast}b + c^{\ast}d) & = & I_{0}D_{xx}\\
I_{0} C_{nl0m} & & = 
               & Im(-a^{\ast}b + c^{\ast}d)& = & I_{0}C_{yz0x} \\
I_{0} K_{m00m} & & = 
               & Re(a^{\ast}c + b^{\ast}d)& = & -I_{0}K_{xx} \\
I_{0} C_{ln0m} & & = 
               & Im(-a^{\ast}c + b^{\ast}d) & = & -I_{0}C_{zy0x}\\
I_{0} C_{nm0m} & & = 
               & Re(b^{\ast}e) - Im(c^{\ast}g) & = & I_{0}C_{yx0x}\\
I_{0} D_{0l0m} & & = 
               &  Re(c^{\ast}g) + Im(b^{\ast}e) & = & I_{0}D_{zx}\\
I_{0} K_{l00m} & & = 
               &  Re(b^{\ast}g) + Im(c^{\ast}e)& = & -I_{0}K_{zx} \\
I_{0} C_{mn0m} & & = 
               &  Re(c^{\ast}e) - Im(b^{\ast}g)& = & -I_{0}C_{xy0x} 
\end{array} \end{equation}
with all other $\cx_{ij00},\cx_{ij0n},\cx_{ij0m} = 0$.  Here we have replaced 
${\cal X}$ by the conventional notation  $P$ for polarization, 
$C$ for spin correlations, $D$ for depolarization, and $K$ for polarization 
transfer. We have included relations to the observables written with the
usual conventions of dropping unnecessary subscripts and utilizing separate 
coordinate axes for the antiparticle and the particle, with $\{\xh,\yh,\zh\}$ 
corresponding to $\{\mh,\nh,\lh\}$ for the $\bar{p}$ and $\bar{\Lambda}$, 
and $\{\xh,\yh,\zh\}$ corresponding to $\{-\mh,\nh,-\lh\}$ for the $p$ 
and $\Lambda$.

Each of these 24 non-zero observables could be independently extracted from 
the angular distribution of the decay baryons.  In such an approach, the 
correction for the experimental acceptance may be problematic, since the 
necessary correction 
may depend on other spin observables.  This approach also ignores the many
non-trivial relationships between the observables~\cite{Richard1,Richard2},
implicit in Eqs. (\ref{relations}), which are a direct result of the 
parity and C-parity symmetries of the reaction. Independent extraction 
of individual observables, or any other procedure that does not embody 
these relationships, does not make optimal use of these symmetries to 
minimize the uncertainty in the extracted parameters, enforce consistency
between the results, or reflect the correlation of their errors.

One way to make full use of the known symmetries would 
be to directly fit the parameters of the transition matrix 
(Eqn.\ \ref{TransMat}).  However, it is not immediately obvious that this 
limited set of observables is sensitive to the entire parameter space. This 
concern is equivalent to the question of whether expressions for the parameters 
$\{a..g\}$ can be found in terms of the accessible observables $\cx_{ij0\nu}$. 
Eqs.\ (\ref{relations}) relate 24 non-trivial observables to 11 real
parameters, with an arbitrary phase. The existence of a consistent 
solution for ideal values of the observables ({\it i.e.} neglecting 
experimental error) is guaranteed by construction.  It then remains 
be to determined whether there is a unique solution or whether there 
exist multiple solutions or even some hyper-surface in parameter 
space which would uniformly satisfy Eqs.\ (\ref{relations}).

A unique solution can be demonstrated to exist over most of the parameter 
space\footnote{For a discussion of the limitations on the range of 
applicability of this demonstration, see the appendix.}. Note that 
the demonstration need not be the optimal algorithm for extraction of the 
parameters with minimal experimental error. The existence of any solution 
implies that the parameters are uniquely determined by this set of 
observables. This is equivalent to the statement that this limited set of
observables completely determines the spin structure of the reaction.
The phase of either $b$ or $c$ can be arbitrarily chosen, then 
both parameters are uniquely determined from the relationships:
\begin{eqnarray}\label{BnC}
|b|^{2} & = & \frac{I_{0}}{2}( 1 - C_{nn00} - K_{n00n} + D_{0n0n} ) \\
|c|^{2} & = & \frac{I_{0}}{2}( 1 - C_{nn00} + K_{n00n} - D_{0n0n} ) \\
b^{\ast}c & = & \frac{I_{0}}{2}( C_{ll00} + C_{mm00} ) + 
           \frac{{\it i}I_{0}}{2}( C_{lm0n} - C_{ml0n} ) 
\end{eqnarray}
Then $a$ and $d$ are uniquely determined from the complex equations:
\begin{eqnarray}\label{AnD}
a^{\ast}b + d^{\ast}c & = &   I_{0}D_{0m0m} - {\it i} I_{0}C_{nl0m} \\
a^{\ast}c + d^{\ast}b & = &   I_{0}K_{m00m} - {\it i} I_{0}C_{ln0m}
\label{AnD2}\end{eqnarray}
which are effectively four linear equations in four unknowns.  Similarly, $e$ 
and $g$ are determined by:
\begin{eqnarray}\label{EnG}
e^{\ast}b - {\it i}g^{\ast}c & = & I_{0}C_{nm0m} - {\it i} I_{0}D_{0l0m} \\
e^{\ast}c - {\it i}g^{\ast}b & = & I_{0}C_{mn0m} - {\it i} I_{0}K_{l00m} 
\label{EnG2}\end{eqnarray}

The 24 observables over-determine the parameter space, so this 
solution is not expressed uniquely.  The existence of a solution implies that 
a global fit of the angular distribution could in principle be done to uniquely
determine the transition matrix parameters.  This implies a complete 
knowledge of the spin structure of the reaction, allowing entirely model 
independent prediction of all spin observables $\cx_{ij\mu\nu}$,
even those whose direct observation would require a longitudinal target 
polarization or the addition of initial antiproton polarization.  In such
an approach, separate fits would be required for each interval in the 
production angle $\theta$ and for each energy point studied.   

An un-binned maximum likelihood fit~\cite{Orear} could be 
applied in order to extract the parameters from the event angular distribution.
This method makes efficient use of statistics and provides a natural way of 
including the detector acceptance through a realistic Monte Carlo simulation.  
Early studies have indicated that such a fit can be successfully performed 
given data sets which are reasonably achievable.

This paper demonstrates that an experiment utilizing a transverse target 
polarization and measuring the angular distribution and final state 
polarizations and correlations of $\reac$ can, in principle, completely 
determine the spin structure of the reaction. This is more information
than was previously thought to be available.
The PS185/3 collaboration has performed such a measurement for the explicit 
purpose of extracting $D_{0n0n}$. One of the authors (kdp) is currently 
active in the analysis of this data and will investigate the possibility of 
extending the extracted results to include the full spin transition matrix 
parameterization as outlined here.  While preliminary indications are 
encouraging, it remains to be seen whether the 
available  statistics are sufficient to yield statistically significant 
results for this extension beyond the original goals of the experiment.

\acknowledgments
This work was supported in part by the U.S. Department of Energy, grant
DE-FG02-87ER40315

\appendix
\section*{}\label{othersoln}

The demonstration shown in Eqs.\ (\ref{BnC}-\ref{EnG2}) of the existence of a 
unique solution for Eqs.\ (\ref{relations}) is not applicable for the special 
case of $b=c$.  With this condition, Eqs.\ (\ref{AnD},\ref{AnD2}) and 
(\ref{EnG},\ref{EnG2}) reduce to only 1 complex equation each, which 
is not sufficient to solve for $\{a,e,d,g\}$.  However, even in the case
of $b=c$ precisely\footnote{This is the condition for the singlet fraction,
$S_{F} = 1 - C_{mm00} - C_{nn00} - C_{ll00} = |b-c|^{2}$, to be 
zero~\cite{Richard1}.  Previous measurements of $\reac$ indicate that $S_{F}$
is very small or zero for $\theta < 90\deg$ and incident 
$\bar{p}$ momentum in the lab frame of less than 1800 MeV/c~\cite{PS185_96}.}, 
a unique solution can still be found. Explicitly taking $b=c$, with $b$ 
chosen to be positive real and known from Eqn.\ (\ref{BnC}), $\{a,g\}$ can 
be determined from the set of 8 linear equations with 8 unknowns:
\begin{equation} \label{soln2} \hspace{-1.0cm}\begin{array}{rcl}
                                  Re(e) - Im(g) & = & F_{1} \\
                                  Im(e) + Re(g) & = & F_{2} \\
                                  Re(d) + Re(a) & = & F_{3} \\
                                  Im(d) + Im(a) & = & F_{4} \\
 F_{1}Re(g) +F_{2}Im(g) -F_{4}Re(a) +F_{3}Im(a) & = & C_{ml0n} \\
-F_{4}Re(g) +F_{3}Im(g) -F_{1}Re(a) -F_{2}Im(a) & = & 
   C_{mm0n} - F_{1}F_{3} - F_{2}F_{4} \\
-F_{4}Re(g) +F_{3}Im(g) +F_{1}Re(a) +F_{2}Im(a) & = & A_{000n} \\
-F_{2}Re(g) +F_{1}Im(g) +F_{3}Re(a) +F_{4}Im(a) & = & 
\begin{array}[t]{r} \frac{1}{2}\left( D_{0n0n} + K_{n00n} -F_{1}^{2}\:\; \right.\\
                      \left. -F_{2}^{2} +F_{3}^{2} +F_{4}^{2} \:\right) \end{array} 
\end{array}\end{equation}
with
\begin{equation} F_{1} = \frac{C_{mn0m}}{b} ,  F_{2} = \frac{D_{0l0m}}{b} , 
 F_{3} = \frac{D_{0m0m}}{b} ,  F_{4} = \frac{C_{ln0m}}{b} \end{equation}

This implies sensitivity to the full set of the scattering matrix 
parameters, even in the case that $b$ and $c$ are equivalent. 
This solution is limited only by the requirement that $b\neq0$.  
In the very special case of $b=c=0$, most of the observables in 
Eqs.\ (\ref{relations}) would be constrained to be zero and it would not be
possible to uniquely determine the full set of parameters using 
only a polarized target. In fact, with $b=c=0$, an experiment involving 
polarizations of both particles in the initial state would be necessary 
in order to be sensitive to the entire parameter space.

An indication of this condition can be found by measuring $C_{nn00}$.  
The relation 
\begin{equation} 1-C_{nn00} = |b|^{2} +|c|^{2} \end{equation}
shows that as $C_{nn00}\rightarrow 1$, both $b$ and $c$ must approach
zero~\cite{Richard1}.  If $C_{nn00} \neq 1$, then one or both of $b$ or $c$ 
must be non-zero. Thus a preliminary study of this observable can be used to 
warn of potential difficulties in performing a global fit to the full 
parameter set.

It should be noted that $C_{nn00}$ has been previously measured for $\reac$
near threshold and that it tends toward the value +1 for only a narrow range 
of production angle $\theta$~\cite{PS185_96}.  This indicates that, with the 
possible exception of that narrow angular range,  determination of the transition 
matrix parameters should, in principle, be possible with data of the type 
collected by PS185/3.



\begin{thebibliography}{16}
%
\bibitem{PS185M}
The PS185 collaboration:
P.D. Barnes {\sl et al.}, Phys. Lett. {\bf B189} (1987) 249.\\
P.D. Barnes {\sl et al.}, Phys. Lett. {\bf B199} (1987) 147.\\
P.D. Barnes {\sl et al.}, Phys. Lett. {\bf B229} (1989) 432.\\
P.D. Barnes {\sl et al.}, Phys. Lett. {\bf B246} (1990) 273.\\
P.D. Barnes {\sl et al.}, Nucl. Phys. {\bf A526} (1991) 575.\\
P.D. Barnes {\sl et al.}, Phys. Lett. {\bf B309} (1993) 469.\\
P.D. Barnes {\sl et al.}, Phys. Lett. {\bf B331} (1994) 203.\\
P.D. Barnes {\sl et al.}, Nucl. Phys. B.(Proc. Suppl.) {\bf 56A} (1997) 46-51.
%
\bibitem{PS185_96}
The PS185 collaboration:\\
P.D. Barnes {\sl et al.}, Phys. Rev. C {\bf 54} (1996) 1877. \\
P.D. Barnes {\sl et al.}, Phys. Rev. C {\bf 54} (1996) 2831. 
%
\bibitem{PS185_3}
The PS185 collaboration: B. Bassalleck {\sl et al}, CERN/SPSLC-95-13/P287 
(3 March 1995)
%
\bibitem{Haidenbauer}
J. Haidenbauer, K. Holinde, V. Mull and J. Speth, Phys. Lett. {\bf
   B291} (1992) 223. \\
J. Haidenbauer, K. Holinde, V. Mull and J. Speth, Phys. Rev. C {\bf 
   46} (1992) 2158.
%
\bibitem{Meson}
F. Tabakin and R. A. Eisenstein, Phys. Rev. C {\bf 31}, 1857 (1985).\\
P. LaFrance and B. Loiseau, Nucl. Phys. {\bf A528} (1991) 557.\\
R.G.E. Timmermans, Th. A. Rijken, and J.J. de Swart, Phys. Rev. D 
{\bf 45} (1992) 2288.
%
\bibitem{Alberg93}
M.A. Alberg, E.M. Henley, P.D. Kunz, and L. Wilets, Nucl. Phys. {\bf A560} 
(1993) 365-388. \\
M.A. Alberg, E.M. Henley, P.D. Kunz, and L. Wilets, Phys. At. Nucl. {\bf 57} 
(1994) 1608-1613.
113.
%
\bibitem{Quark}
S. Furui and A. Faessler, Nucl. Phys. {\bf A468} (1987) 669.\\
M. Kohno and W. Weise, Phys. Lett. {\bf B206} 584 (1988).
%
\bibitem{Alberg95}
M.A. Alberg, J. Ellis and D. Kharzeev, Phys. Lett. {\bf B356} (1995) 
113.
%
\bibitem{Pak99}
N.K. Pak and M.P. Rekalo, Phys. Lett. {\bf B450} (1999) 433.
%
v\bibitem{Wolfenstein}
L. Wolfenstein and J. Ashkin, Phys. Rev. {\bf 85} (1952) 947.
%
\bibitem{Bystricky78}
J. Bystricky, F. Lehar and P. Winternitz, J. de Physique
  (Paris) {\bf 39} (1978) 1.
%
\bibitem{LaFrance80}
P. La France and P. Winternitz, J. de Physique (Paris) {\bf 41} (1980) 
1391.
%
\bibitem{Durand}
L.Durand III and J. Sandweiss, Phys. Rev. {\bf 135} (1964) B540
%
\bibitem{Richard1}
J.-M. Richard, Phys. Lett. {\bf B369} (1996) 358.
%
\bibitem{Richard2}
M.Elchikh and J.-M. Richard, Phys. Rev. C {\bf 61}(2000)035205
%
\bibitem{Orear}
J. Orear, ``Notes on Statistics for Physicists, Revised'', Cornell
Laboratory Nuclear Studies, CLNS 82/511 (July 28, 1982).
%
\end{thebibliography}
\end{document}